\documentclass[aps,preprint,amsmath,amssymb]{revtex4}
\usepackage{graphicx}
\usepackage{color}

\def\g5{\gamma_{5}}
\def\ga{\gamma}
\def\ve{\varepsilon}

\def\be{\begin{eqnarray}}
\def\ed{\end{eqnarray}}

\begin{document}
\title{\Large \bf Non-standard neutrino interactions in\\
$K^{+}\to\pi^{+}\nu\bar{\nu}$ and $D^{+}\to\pi^{+}\nu\bar{\nu}$ decays }
\date{\today}

\author{ \bf  Chuan-Hung Chen$^{1,2}$\footnote{Email:
{\sf physchen@mail.ncku.edu.tw}}, Chao-Qiang
Geng$^{3,4}$\footnote{Email: {\sf geng@phys.nthu.edu.tw}},
and
 Tzu-Chiang Yuan$^{3}$\footnote{Email: {\sf tcyuan@phys.nthu.edu.tw}}
 }

\affiliation{ $^{1}$Department of Physics, National Cheng-Kung
University, Tainan 701, Taiwan \\
$^{2}$National Center for Theoretical Sciences, Taiwan\\
$^{3}$Department of Physics, National Tsing-Hua University, Hsinchu
300, Taiwan  \\
$^{4}$Theory Group, TRIUMF, 4004 Wesbrook Mall\\Vancouver, B.C V6T 2A3, Canada
 }

\begin{abstract}
We study the contributions of the non-standard neutrino interactions
(NSIs) to the rare meson decays of $K^{+}\to \pi^{+}  \nu\bar{\nu}$
and $D^{+}\to \pi^{+} \nu\bar\nu$. We show that both decays could
provide strong constraints on the free parameters in the NSIs. We
point out that the branching ratio of $D^{+}\to \pi^{+} \nu\bar\nu$
could be at the level of $10^{-8}$.
\end{abstract}
\maketitle
It has been widely anticipated to have new physics
beyond the Standard Model (SM) in the neutrino sector
due to the results of atmospheric and solar neutrino oscillations experiments \cite{PDG06}.
In particular, it has been proposed to have
new effective four-fermion interactions involving neutrinos, called
non-standard neutrino interactions (NSIs) \cite{NSI1,NSI2,NSI3,NSI4},
given by \cite{np3}
 \be
{\cal L}^{\rm NSI}_{\rm eff}&=& - 2\sqrt{2} G_F \ve^{fP}_{\ell \ell'}
(\bar \nu_{\ell} \ga_{\mu} L \nu_{\ell'} )
(\bar f \ga^{\mu} P f ) \,,
\label{eq:nsi}
 \ed
where the index of $\ell (\ell')$ corresponds to the light neutrino flavor,
$f$ denotes a charged lepton or quark in the first
generation, and $P=L$ or $R$ with $L(R)=(1\mp \ga_5)/2$. It has been
shown that the NSIs would be compatible with the oscillation effects
along with some new features in various neutrino searches
\cite{np1,np2,KSY}. The experimental constraints on NSIs
have been studied in Refs. \cite{np3,constraint1,np4}.
%
In particular, the
authors in Ref.~\cite{np3} have also pointed out that the lepton flavor violating decays of $\mu\to 3
e$, $\tau\to (e,\, \mu) ee$, $\tau \to (e,\, \mu)M$ with $M=\rho$
or $\pi$, and the $\mu-e$ conversion on nuclei
could be induced through one-loop effects and give
some stringent constraints on the parameters of $\ve^{fP}_{\ell \ell'}$ in Eq. (\ref{eq:nsi}).
Explicitly,
$\ve^{fP}_{\mu e}$ and $\ve^{fP}_{\tau \ell}\ (\ell=e,\mu)$
are limited to be less than $10^{-3}$ and order of one, respectively.
On the other hand, for the flavor diagonal NSIs in the first two generations,
$\ve^{fP}_{ee,\mu\mu}$ are constrained by the tree level processes in the low energy scattering experiments and could be limited at the level of $O(10^{-3})$ in
future $\sin^{2}\theta_{W}$ experiments.
For the third generation, $\ve^{fP}_{\tau\tau}$ is presently bounded to be order of one
by the LEP experiments but its future limit can be slightly improved to be
$O(0.3)$ \cite{np3}
from KamLAND \cite{KL} and solar neutrino data \cite{SNO,SK}.
Note that although the constraints on the parameters of
$\ve^{fP}_{\tau \ell}\ (\ell=e,\mu)$
can be given by the precision measurements on
$\sin^2\theta_{W}$ at neutrino factories but the bounds can only be
$O( 10^{-2})$ \cite{np3,nuf}.

In this brief report, we will study two new processes of
$K^{+}\to \pi^{+} \nu\bar\nu$ and
$D^{+}\to \pi^{+} \nu\bar\nu$
in the framework of the NSIs. We will show that the
parameters involving $\tau$ neutrino in $\ve^{uL}_{\ell \ell'}$
could be limited to be less than $10^{-2}$ by the rare decay
$K^{+}\to \pi^{+}\nu\bar\nu$.
We will also display
the branching ratio (BR) of
$D^{+}\to \pi^{+} \nu_{\ell} \bar{\nu}_{\ell'}$ could be
at the level of $10^{-8}$ due to the NSI, which could
be accessible to
BESIII \cite{bes3}.

We start with the decay of $K^{+}\to \pi^{+} \nu\bar \nu$. Since we
are not going to deal with CP violation, we will skip the discussion
of the CP violating mode of $K_{L}\to \pi^{0} \nu\bar\nu$, which is
nevertheless interesting on its own \cite{GIM_PLB}. It is well known
that at the quark level,
 the effective Hamiltonian for $s\to d \nu
\bar \nu$ in the SM is given by \cite{BBL}
\be H&=& \frac{G_{F}}{\sqrt{2}} \frac{\alpha_{em}}{2\pi
\sin^2\theta_{W}} \sum_{\ell=e,\mu,\tau}\left(
V^{*}_{cs}V_{cd}X^{\ell}_{NL}+V^{*}_{ts}V_{td}X(x_t)\right)(\bar s
d)_{V-A}(\bar\nu _{\ell} \nu_{\ell})_{V-A}
 \ed
 where $(\bar f f')_{V-A}=\bar f \gamma_{\mu}(1-\gamma_5) f'$, $X^{\ell}_{NL}$ denotes
 the charm quark contributions,
and $X(x_t)$ is the loop integral of the top quark contribution given by
\begin{equation}
X(x_t)=
\eta_{X}\frac{x_t}{8}\left[\frac{x_t+2}{x_t-1}+\frac{3x_t-6}{(x_t-1)^2}
\ln x_t \right]
\end{equation}
with $x_t=m^2_{t}/m^2_{W}$ and $\eta_{X}=0.985$ being the QCD
short-distance correction.
The SM prediction for ${\rm
BR}(K^{+}\to \pi^{+}\bar \nu \nu)_{\rm SM}$ at the
next-to-next-to-leading order is found to be $(8.0\pm 1.1)\times
10^{-11}$ \cite{buras}, which is
 smaller than the data of ${\rm BR}(K^{+}\to \pi^{+}
\nu\bar\nu)_{\rm exp}=(1.5^{+1.3}_{-0.9})\times 10^{-10}$ \cite{E949}.
If future
data still keep the tendency toward a larger central value, due to the
small theoretical uncertainties \cite{LD},
it should indicate the existence of new physics.
Note that with ${\rm BR}(K^{+}\to \pi^{+}
\nu \bar\nu)\sim 10^{-10}$ it is possible to collect 40 signal events per
year at the NA48/3 experiment of CERN-SPS \cite{NA48/3,P-326}.

\begin{figure}[htbp]
\includegraphics*[width=2.5 in]{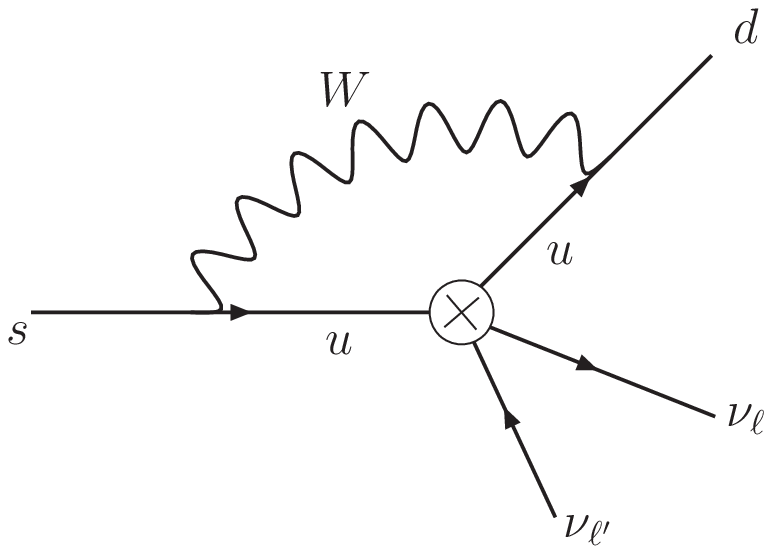}
\caption{Flavor diagram for the $s\to  d \nu_{\ell}
\bar{\nu}_{\ell'}$.}
 \label{fig:nsi}
\end{figure}

The NSIs in Eq.~(\ref{eq:nsi}) can also induce the process
$s\to d \nu_{\ell} \bar{\nu}_{\ell'}$ through a one-loop diagram,
as illustrated in Fig.~\ref{fig:nsi}.
As a result, the following four-fermion interaction can be induced
 \be
 \label{H-NSI}
H^{\rm NSI}_{s \to d \nu_{\ell}\bar{\nu}_{\ell'} }&=& - \frac{G_F}{\sqrt{2}} \left(
V^{*}_{us} V_{ud} \frac{\alpha_{em} }{4\pi
\sin^2\theta_W}\ve^{uL}_{\ell \ell'}\ln\frac{\Lambda}{m_W}\right)
(\bar {\nu}_{\ell}  \nu_{\ell'})_{V-A} (\bar s  d)_{V-A}
 \ed
 where $\Lambda$ denotes the new physics
 energy scale above the weak scale
 with $\ln(\Lambda/m_W)>1$.
%
 From Eq. (\ref{H-NSI}),
the BR for $K^+ \to \pi^+ \nu_{\ell}
\bar{\nu}_{\ell'}$ from the NSIs is found to be
 \be
{\rm BR}(K^{+}\to \pi^{+} \nu_{\ell}
\bar{\nu}_{\ell'})_{\rm NSI}
&=&
\frac{\kappa_{+}}{3} \left| V^{*}_{us}V_{ud} \frac{1}{2}
\ve^{uL}_{\ell \ell'} \ln\frac{\Lambda}{m_W}\right|^2{\rm BR}(K^{+}\to \pi^{0} e^{+} \nu_{e})\,,
\label{eq:brkpinunu2}
 \ed
 with
  \begin{eqnarray}
 &&\kappa_{+}=r_{K^+} \frac{3\alpha^2_{em}}{|V_{us}|^2 2\pi^2
 \sin^4\theta_{W}}\,,
 \end{eqnarray}
 where   the factor of 3  in $\kappa_{+}$ comes
 from the number of neutrino species and $r_{K^{+}}=0.901$ denotes
 the isospin breaking effects \cite{MP}.
 Numerically, it is given by
 \be
 \label{eq:num}
  {\rm BR}(K^{+}\to \pi^{+} \nu_{\ell} \bar{\nu}_{\ell'})_{\rm NSI}
\approx 6.5 \times 10^{-7}
\left\vert \ve^{uL}_{\ell \ell'}
\ln\frac{\Lambda}{m_W}\right\vert ^2\,,
 \ed
 where we have used
 ${\rm BR}(K^{+}\to \pi^{0} e^{+} \nu_{e})=(4.98\pm 0.07)\%$
\cite{PDG06}.
For simplicity, we will avoid the complicated interference effects
to get the constraint on the free parameters directly by setting
the upper bound for the contribution from the NSIs in Eq. (\ref{eq:num}) to be less than
 ${\rm \Delta BR \equiv BR_{exp}-BR_{SM}} = {\cal C}\times 10^{-10}$ with ${\cal C}\sim 0.5$.
Hence, we obtain the upper limit on $\ve^{uL}_{\ell
\ell'}$ to be
 \be
  \ve^{uL}_{\ell \ell'} \lesssim \frac{{\cal C}}{0.5} \frac{8.8\times
  10^{-3}}{\ln\Lambda/m_W}\,.
 \ed
Since $\ell$ or $\ell'$ can represent any charged lepton,
the decay of $K^+\to \pi^+\nu\bar\nu$ could give  strong
constraints on the parameters involving $\nu_{\tau}$.

Next, we discuss the NSIs on $D^{+}\to \pi^{+} \nu\bar\nu$. It is well known that
unlike the $K$ and $B$ systems, where the
FCNC processes are enhanced by the internal heavy top-quark loop,
the loop induced rare decays of the charmed mesons are highly suppressed in the
SM as they involve the light down-quark sector  and
the GIM cancellation.
 It has been
estimated that
${\rm BR}(D^{+} \to \pi^{+} \nu\bar \nu)_{\rm SM}\sim 5.1\times 10^{-16}$  with
long-distance QCD contributions included  in the SM \cite{Ddecay}.
Obviously, it cannot be reached by future experiments such
as BESIII,
where the sensitivity on the rare charmed
meson decays
will be reached only at the level of $10^{-8}$ or so \cite{bes3}.
It is clear that if there is any evidence with
${\rm BR}(D^{+}\to \pi^{+} \nu\bar\nu)\sim 10^{-8}$ in the future experiment, it will
definitely indicate the existence of new physics.

\begin{figure}[htbp]
\includegraphics*[width=2.5in]{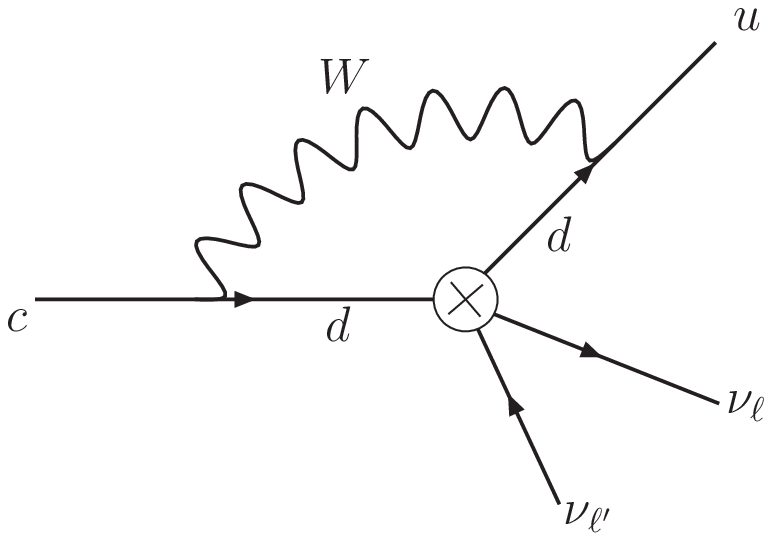}
\caption{Flavor diagram for the $c\to  u\nu_{\ell}
\bar{\nu}_{\ell'}$.}
 \label{fig:nsi2}
\end{figure}
The NSIs in Eq.~(\ref{eq:nsi}) can also induce the transition $c\to
u\nu_{\ell} \bar{\nu}_{\ell'}$ at one-loop level with a Feynman diagram displayed in
Fig.~\ref{fig:nsi2}. Similar to $s\to d
\nu_{\ell}\bar{\nu}_{\ell'}$, the following effective interaction for $c\to
u\nu_{\ell} \bar{\nu}_{\ell'}$ can be obtained
 \be
 H^{\rm NSI}_{c \to u\nu_{\ell}\bar{\nu}_{\ell'}}&=& \frac{G_F}{\sqrt{2}}  \left( \frac{\alpha_{em}}{4\pi \sin^2\theta_W}
 V_{cd} V^{*}_{ud} \ve^{dL}_{\ell \ell'} \ln \frac{\Lambda}{m_W} \right)
 (\bar \nu_{\ell} \nu_{\ell'})_{V-A} (\bar c u)_{V-A}\,,
 \ed
which leads to
 \be
 {\rm BR}(D^{+}\to \pi^{+} \nu_{\ell} \bar{\nu}_{\ell'})_{\rm NSI}&=& \left\vert
   V^{*}_{ud}\frac{\alpha_{em}}{4\pi\sin^2\theta_W}  \ve^{dL}_{\ell \ell'}
 \ln\frac{\Lambda}{m_W} \right\vert^2 {\rm BR}(\bar D^{0}\to \pi^{+} e \bar
 \nu_{e})\,.
 \ed
Numerically, we obtain
  \be
   {\rm BR}(D^{+}\to \pi^{+} \nu_{\ell} \bar{\nu}_{\ell'})_{\rm NSI}\approx 2\times
   10^{-8} \left\vert \ve^{dL}_{\ell\ell'} \ln\frac{\Lambda}{m_W}
   \right\vert ^2\,,
  \ed
where we have used ${\rm BR}(\bar D^{0}\to \pi^{+} e \bar
\nu_{e})=(2.81\pm 0.19)\times 10^{-3}$ \cite{PDG06}.
For $\ln(\Lambda/m_W)\sim 1$,
$\ve^{dL}_{\tau \tau}\sim 1$, and $\ve^{dL}_{\ell\ell'}<1\
(\ell,\ell'\neq \tau)$, we get ${\rm BR}(D^{+}\to \pi^{+}
\nu\bar{\nu})\sim 2\times 10^{-8}$,
which could be reached at a future dedicated
experiment \cite{BG} such as
BESIII \cite{bes3}.
Turning this argument around, if future searches of these rare $D$ decay modes
at the level of $10^{-8}$ turn out to be nil at BESIII, useful constraints on
the couplings $\ve^{dL}_{\ell\ell'}$ can be deduced.

To recap, we have studied $K^{+}\to \pi^{+} \nu\bar\nu$ and
$D^{+}\to \pi^{+} \nu\bar\nu$ in the framework of the NSIs. We have
shown that both rare decays could provide strong constraints on the free
parameters in the NSIs. Explicitly, we have found that
$\ve^{uL}_{\ell \ell'}$ and $\ve^{dL}_{\ell\ell'}$ could be limited
to be less than $10^{-2}$ and order of one by the current and future
experiments for $K^{+}\to \pi^{+}\nu\bar\nu$ and $D^{+}\to \pi^{+}
\nu\bar\nu$, respectively. Finally, we remark that ${\rm
BR}(D^{+}\to \pi^{+}\nu \bar\nu)$ could be at the level of $10^{-8}$
due to the NSIs, which could be
accessible to BESIII \cite{bes3}.
\\


 \noindent {\bf Acknowledgments}

We would like to thank Prof. D. Bryman and Dr. Hiroaki Sugiyama for useful
discussions. This work is supported in part by the National Science
Council of R.O.C. under Grant \#s: NSC-95-2112-M-006-013-MY2,
 NSC-95-2112-M-007-001, NSC-95-2112-M-007-059-MY3, and
 NSC-96-2918-I-007-010.

\end{document}